# Maunakea Spectroscopic Explorer Low Moderate Resolution Spectrograph Conceptual Design

Patrick Caillier[a], Will Saunders[b], Pierre-Henri Carton[c], Florence Laurent[a], Jean-Emmanuel Migniau[a], Arlette Pécontal[a], Johan Richard[a], Christophe Yèche[c]

[a]Université de Lyon, Lyon, F-69003, France; Université Lyon 1, Observatoire de Lyon, 9 avenue Charles André, Saint-Genis Laval, F-69230, France; CNRS, UMR 5574, Centre de Recherche Astrophysique de Lyon; Ecole Normale Supérieure de Lyon, Lyon, F-69007, France; [b]Australian Astronomical Observatory, Camberra, Australia; [c]Commissariat à l'énergie atomique et aux énergies alternatives

## ABSTRACT

The Maunakea Spectroscopic Explorer (MSE) Project is a planned replacement for the existing 3.6-m Canada France Hawaii Telescope (CFHT) into a 10-m class dedicated wide field highly multiplexed fibre fed spectroscopic facility. MSE seeks to tackle basic science questions ranging from the origin of stars and stellar systems, Galaxy archaeology at early times, galaxy evolution across cosmic time, to cosmology and the nature of dark matter and dark energy. MSE will be a primary follow-up facility for many key future photometric and astrometric surveys, as well as a major component in the study of the multi-wavelength Universe.

The MSE is based on a prime focus telescope concept which illuminate 3200 fibres or more. These fibres are feeding a Low Moderate Resolution (LMR) spectrograph and a High Resolution (HR). The LMR will provide 2 resolution modes at R>2500 and R>5000 on a wavelength range of 360 to 950 nm and a resolution of R>3000 on the 950 nm to 1300 nm bandwidth. Possibly the H band will be also covered by a second NIR mode from ranging from 1450 to 1780 nm. The HR will have a resolution of R>39000 on the 360 to 600 nm wavelength range and R>20000 on the 600 to 900 nm bandwith.

This paper presents the LMR design after its Conceptual Design Review held in June 2017. It focuses on the general concept, optical and mechanical design of the instrument. It describes the associated preliminary expected performances especially concerning optical and thermal performances.

**Keywords:** MSE, CFH, CFHT, Spectrograph, multi-object, design, performances

## 1 INTRODUCTION

This paper is the Conceptual Design Report for the Low-Moderate Resolution Spectrograph Sub-System (LMR) of Maunakea Spectroscopic Explorer (MSE) and intends to describe the design proposed for the LMR and show how it will comply with its main requirements.

These requirements are given in the Low Resolution Spectrograph Requirements Document which itself carries on the specification work done in the "Conceptual Design Objectives for Low Resolution Spectrograph" and flows down the Science Requirements Documents (SRD) and Observatory Requirement Document (ORD).

# 2 LMR SYSTEM OVERVIEW

## 2.1 Introduction

MSE's coverage requirement at Low Resolution is 360-1300nm. This requires multiple detector technologies, and is also much larger than can efficiently be covered by a single VPH grating. Hence it is obvious that the LMR spectrograph will have multiple arms. The cost is largely driven by the NIR detector, and this is minimised by using a single detector in a fast camera. The largest NIR detectors likely to be available are Hawaii 4RG15's, 61mm x 61mm. Given the fiber size and NIR resolution requirements, the entire *YJ* wavelength range (950-1300nm) can be covered by a single detector, given a sufficiently fast ($\leq$ F/1.2) NIR camera. F/1.2 is a very challenging speed for a transmissive camera, but it offers multiple advantages over a catadioptric design:

(a) Increased throughput because of the unobscured pupil,
(b) Reduced ghosting and scattering for the same reason,
(c) Much easier detector mounting and access,
(d) A more compact design. Hence transmissive cameras are used in this design.

The required number of resolution elements for the Optical range (360-950nm), and the need for dichroic overlaps, means it is not (quite) possible to cover the Optical range with two cameras (assuming square detectors and the slit covering the full width of the detector). In any case, with two cameras, the range for each VPH would be very large, and VPH efficiency very low at the ends of the range of each. Hence we have adopted a 4-armed design, with 3 Optical (Blue, Green, Red) and 1 NIR channel. A 4-armed design is only slightly more difficult to lay out than a three armed design (as adopted for DESI, 4MOST and PFS). Conveniently, there is space around the NIR arm, allowing it to be cooled further than the rest of the spectrograph (as needed for the H-band).

The Optical cameras can be either

(a) The same speed and detector size as the NIR camera, or
(b) Use larger 92mm x 92mm detectors, operating at ~F/1.8.

The latter option means two fundamentally different camera designs, and a significant increase in cost and volume. Hence the all-F/1.2 option was pursued first, to see if adequate image quality could be achieved in all arms.

The required change in resolution is achieved by exchanging the VPH gratings for sapphire grisms containing a higher dispersion VPH. This gives a resolution increase factor of ~1.8 (the refractive index of sapphire).

Finally, the number of fibers to address in total by the LMR is of 3200 to 3468 given the possible fiber focal plane configuration. Given the chip size and f number one LMR unit can accommodate around 550 fibers. 6 LMR units are therefore needed in total.

## 2.2 Optical Design Overview

The design presented here is an evolution of a design laid out for Hector [2], which has similar resolution requirements and optical wavelength range, and the same constraint of trying to minimize the cost per spectrum. The demagnification is 0.58, and the beam size is 175mm. The pupil is not on the grating, but just inside the camera, since this improves image quality and reduces the required lens sizes. The optical layout is shown in Figure 1.

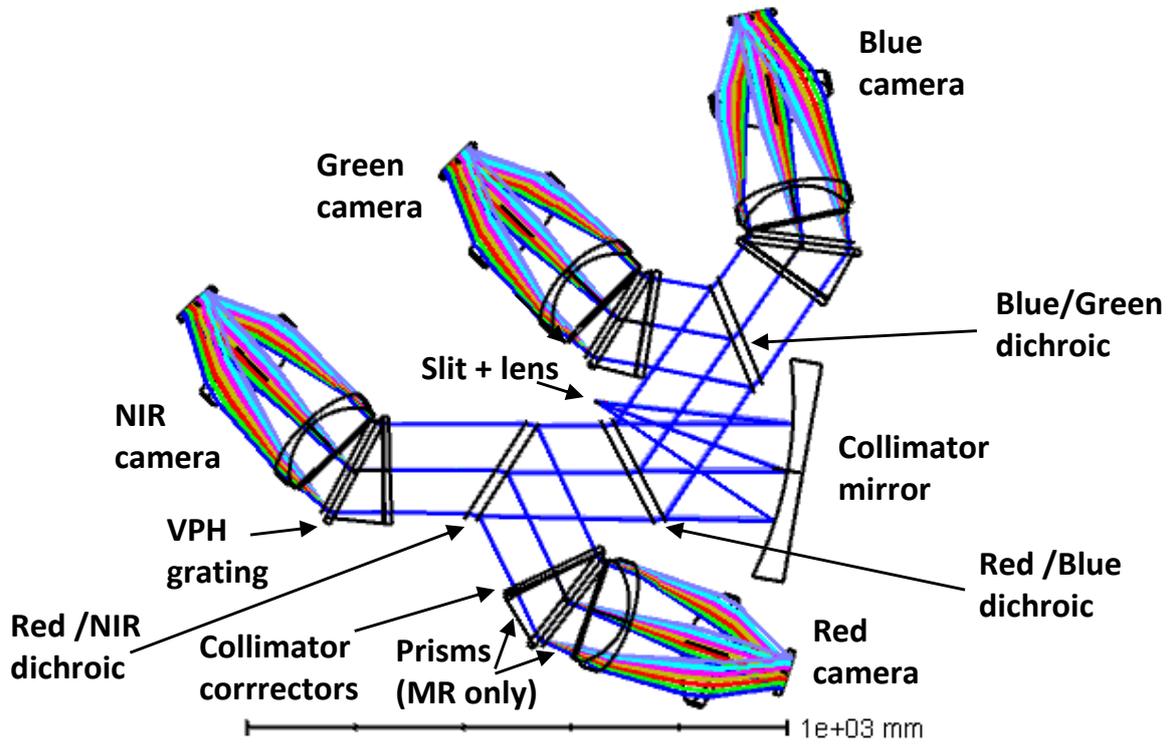

Figure 1. Optical layout for the LMR spectrograph, showing both LR and MR configurations.

The slit is 104.5mm long, and curved laterally (with sag 1.3mm), to straighten its image on the slit. There is a rectangular field-lens gelled to the fibers, as for AAOmega [3]. This increases throughput (since it will be AR coated on the exit side), and allows the spectrograph pupil to be placed as wanted. There is an off-axis Schmidt collimator, with the dichroics between the collimator mirror and the corrector lenses. For LR use, there is an off-axis Schmidt corrector lens for each arm, and then the VPH grating. For MR (and also H-band) use, these are exchanged for a large sapphire grism, with the VPH immersed within it, and with the collimator corrector bonded to the first surface.

There is then a very similar F/1.2 camera for each arm. Each camera contains two glued doublets, with glass choice depending on the camera. For the Blue camera, the doublets consist of i-line glasses, for maximum transmission. The field flattener is AlON (see below), and also acts as dewar window. Each arm has a single 61mm x 61mm detector, either an E2V 231-series CCD, or a Teledyne Hawaii 4RG15. In either case the pixel size is 15µm, and the images are well sampled at ~3.5 pix/FWHM.

## 2.3 Mechanical Design Overview

From the optical design the following mechanical design and set of the sub-assemblies have been defined:

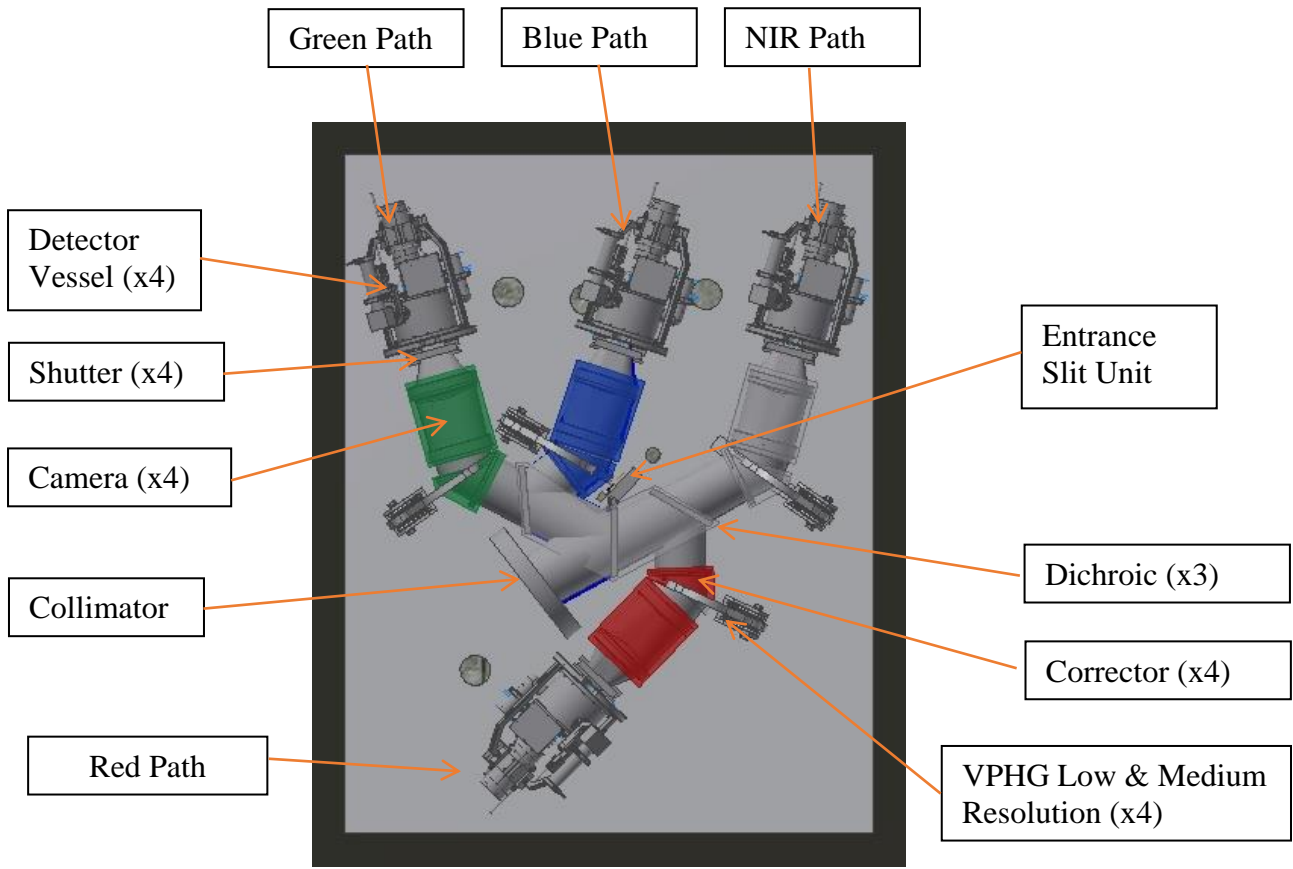

Figure 2: LMR Top View

The main mechanical components in addition to the optical ones are:

- The Optical Bench which ensure optical components fastening and common reference plane for alignment.
- The instrument enclosure or cover assembly enabling dust and environmental protection, proper thermal insulation, light tightness and gas tightness especially for the cooled H band option.
- A support structure to hold the optical table and other miscellaneous components
- Detector Vessels including cryogenics capability with an embedded Linear Pulse Tube (LPT) and associated vacuum system
- Associated Instrument & Detector Control Electronics & Software

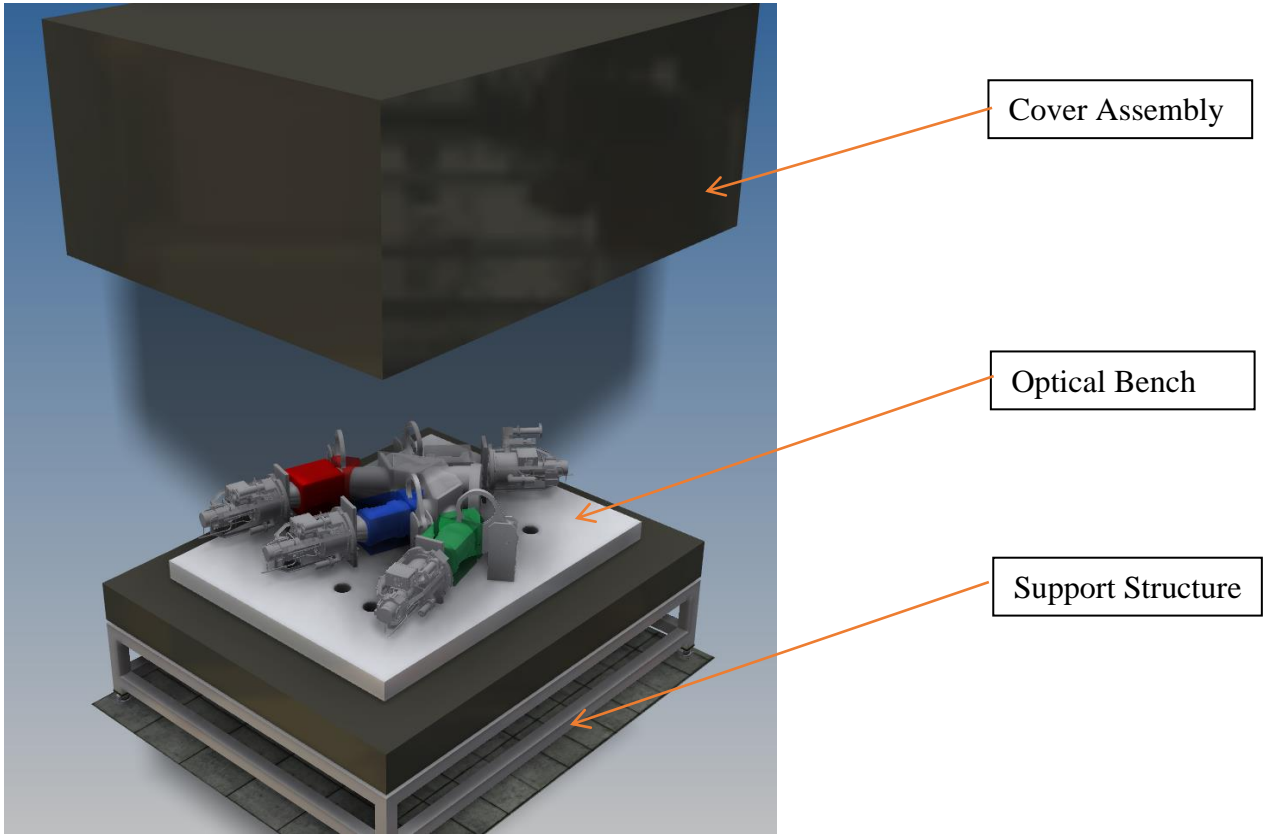

Figure 3: LMR Global View

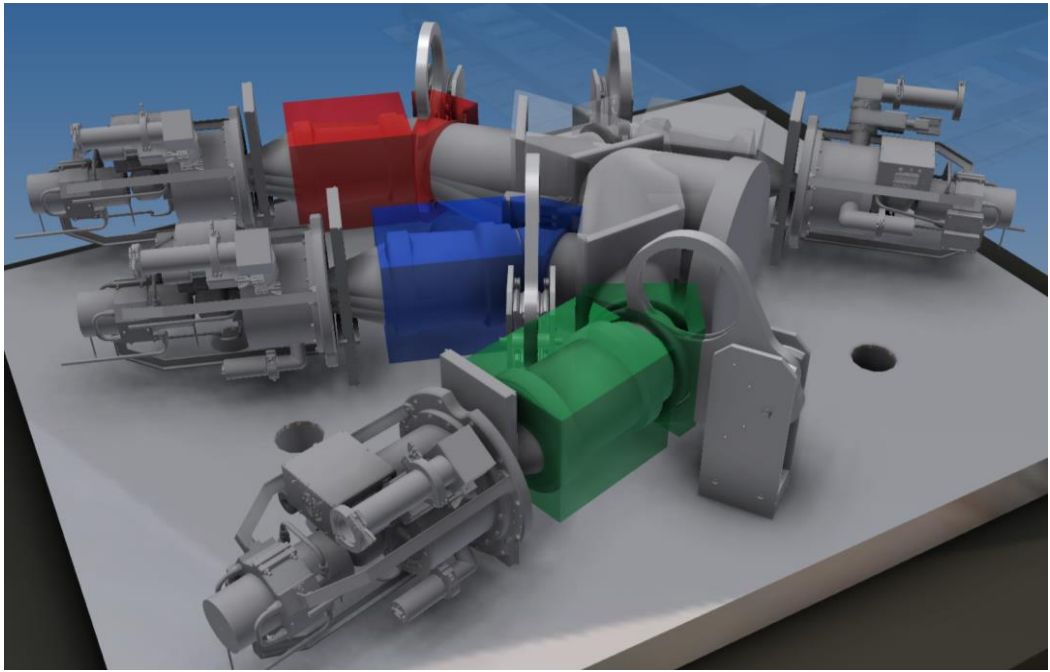

Figure 4: LMR Close-up View

# 3 OPTICAL DESIGN DESCRIPTION

## 3.1 Wavelength coverage and nominal resolution

The wavelength coverage and nominal resolution for each arm is shown in Table 1.

Table 1. Nominal wavelength and resolution parameters for design

|  | LR | | | | MR | | | |
|---|---|---|---|---|---|---|---|---|
|  | **Blue** | **Green** | **Red** | **NIR** | **Blue** | **Green** | **Red** | **NIR** |
| $\lambda_{min}$ | **360nm** | **540nm** | **715nm** | **960nm** | **391nm** | **576nm** | **737nm** | **1457nm** |
| $\lambda_{max}$ | **560nm** | **740nm** | **985nm** | **1320nm** | **510nm** | **700nm** | **900nm** | **1780nm** |
| **Resolution (Å)** | **1.78Å** | **1.75Å** | **2.36Å** | **3.15Å** | **1.02Å** | **1.02Å** | **1.35Å** | **2.68Å** |
| $R_{min}$ | **2000** | **3100** | **3000** | **3000** | **3800** | **5600** | **5500** | **5400** |
| $R_{max}$ | **3100** | **4200** | **4200** | **4200** | **5000** | **6800** | **6700** | **6600** |

These numbers were largely carried over from Hector, where a lower resolution was wanted in the Blue arm. The design has been recently re-optimized in the light of the current requirements, with essentially identical image quality, throughput etc. The revised parameters are shown in Table 2. The resolution requirements are met, except for just missing the minimum MR resolution requirement in the Green and Red arms.

Table 2. Nominal parameters for a revised design with essentially identical image quality and throughput

|  | LR | | | | MR | | | |
|---|---|---|---|---|---|---|---|---|
|  | **Blue** | **Green** | **Red** | **NIR** | **Blue** | **Green** | **Red** | **NIR** |
| $\lambda_{min}$ | **360nm** | **495nm** | **680nm** | **940nm** | **370nm** | **504nm** | **719nm** | **1452nm** |
| $\lambda_{max}$ | **515nm** | **710nm** | **975nm** | **1300nm** | **460nm** | **630nm** | **900nm** | **1780nm** |
| **Resolution (Å)** | **1.36Å** | **1.89Å** | **2.59Å** | **3.14Å** | **0.75Å** | **1.06Å** | **1.52Å** | **2.73Å** |
| $R_{min}$ | **2600** | **2600** | **2600** | **3000** | **5000** | **4800** | **4700** | **5300** |
| $R_{max}$ | **3800** | **3800** | **3900** | **4100** | **6200** | **5900** | **5900** | **6500** |

## 3.2 Collimator speed

The light exits the fibers in a strongly apodised beam (Figure 5 (a)), because of (a) the non-circular M1, (b) Focal Ratio Degradation within the fibers, and (c) (only if the Sphinx positioner is selected) geometric FRD caused by fiber tilt. The demagnification between cameras and collimator is fixed by the fiber size and resolution requirements. So there is a trade-off between camera and collimator speeds, and we have accepted some light loss at the collimator to allow an easier camera design. If the fiber size is reduced, this would allow a faster collimator and hence reduced collimator losses; this trade-off was not included in the selection of the optimal fiber size, but should be done before Preliminary Design is started.

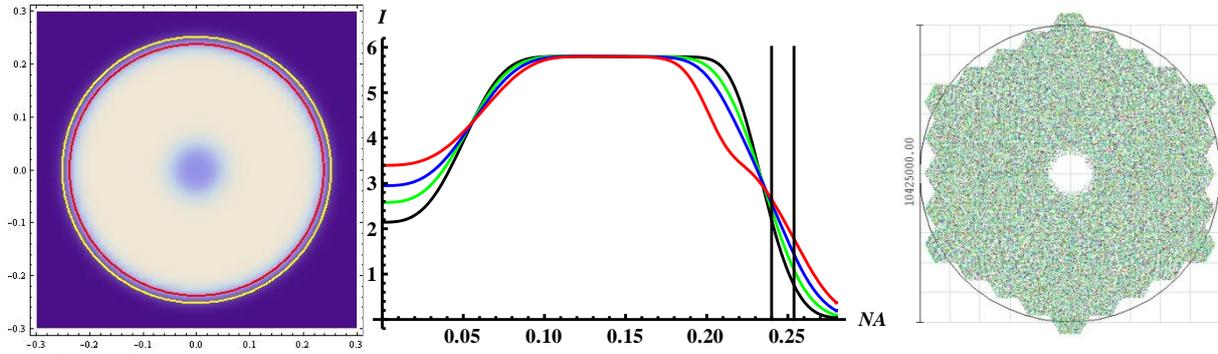

Figure 5 . (a): simulated far-field image of the fiber output for an on-axis untilted fiber, showing collimator acceptance speeds of F/2.083 (red) and F/1.97 (yellow). (b): radial profile through the same beam (black), and also for fibers tilted at 1°, 1.5°, 2°. The collimator speeds of F/2.08 and F/1.97 are also shown. The collimator overfilling light loss for an untilted fiber is 4.5% and 2.2% respectively. (c) M1, with a circle corresponding to F/2.083 superimposed.

The selected collimator speed of F/2.083 corresponds to a diameter of 10.42m at M1 (Figure 5 (c)). The collimator overfilling loss, for a fiber near the center of the field, and assuming untilted fibers and FRD as found by Zhang et al [1], is 4.5%. An F/1.97 collimator would still have losses of 2.2%. These numbers are for central fibers, fibers away from the center of the field would have smaller losses.

### 3.3 Collimator design

An F/2.083 collimator is still very fast. The initial design had an on-axis Schmidt collimator design, but this means the slit has to be buried within the first dichroic, increasing the obstruction losses and making back-illumination more difficult. Amazingly, an unobscured off-axis design works fine, with just a singlet corrector in each arm, with slightly different prescriptions. Being off-axis means the collimator corrector lens is a freeform asphere, but the optical axis would be included within the aperture, making testing and alignment straightforward.

The off-axis design means that the fiber face is not orthogonal to the optical axis, so there is some defocus across the face of the fiber. However, the effect is small, adding 1.25μm to the RMS spot radius, in the spectral direction only.

Also remarkably, an off-axis collimator apparently also allows enough room for the slit, dichroics, dispersers and cameras, though some clearances are small. The off-axis tilt is 10°, large enough to give 30mm of clearance between the beam and the slit, to accommodate a 4MOST-style shutter. However, with a simpler shutter at the slit (and Bonn shutters in each camera), this angle could be reduced. This would also improve the image quality, both via the reduced speed of the parent Schmidt corrector lens, and via the reduced defocus across the fiber.

### 3.4 Dichroics

The dichroics are used at angles of 27.5° (the first dichroic) and 32.5° (the other two). The width of the cross-over regime is strongly dependent on this angle (Figure 6). There is also a small (up to 0.5%) shift in dichroic wavelength with spatial field angle. The crossover width used in the design is ~4%, which should cover the 10%/90% transmission/reflection width for all fibers.

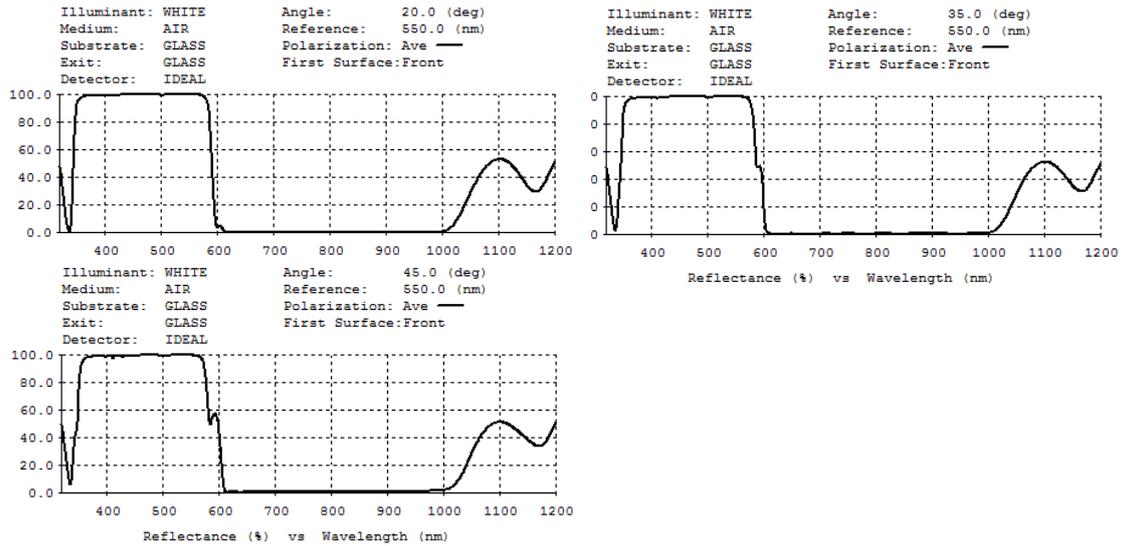

Figure 6. Predicted dichroic responses from Crystal Optics, for 20°, 35°, 45° AOI, all for (370nm-580nm) + (580nm-1000nm). 10%/90% widths are ~12nm, 21nm, 31nm respectively.

The first dichroic at 700nm is challenging, because both the transmitted (700-1800nm) and reflected (360-700nm) wavelength ranges are so large. The other two dichroics are straightforward.

### 3.5 Gratings and grisms

The gratings are straightforward VPH gratings of reasonable (200mm) size, angles, line densities, DCG thickness and index modulation. The grating fringes are slanted at 4.5° (i.e. $\alpha - \beta = 9°$) to throw the Littrow ghost [3] off the detector entirely; the resulting anamorphism allows a useful resolution gain, and also compensates for the increase in beam size in the spectral direction caused by the pupil being below the grating.

Table 3. Gratings characteristics

|  | Blue | Green | Red | NIR |
|---|---|---|---|---|
| **LR lines/mm** | 1390 | 1332 | 988 | 743 |
| **MR lines/mm** | 2275 | 2130 | 1643 | 825 |
| **$\alpha$ (in air for LR, in sapphire for MR)** | 21.7° | 27.9° | 27.6° | 27.9° |
| **$\beta$ (in air for LR, in sapphire for MR)** | 12.7° | 18.9° | 18.6° | 18.9° |

The LR VPH gratings are mounted on fused silica substrates. This gives excellent thermal stability (CTE=0.5ppm/°C). The prisms used in the grisms are of sapphire, and these are large and heavy (8kg). The birefringence has no effect in the collimated beam. They give an increase in resolution of a factor 1.75 over LR use. To achieve a larger resolution increase would require putting reverse prisms on the LR gratings, which would push up the required clearances on both sides of the grating, increasing lens sizes and decreasing image quality. There is a further asphere on the exit surface of the grism, giving a small improvement in image quality.

### 3.6 Camera barrel

Each camera barrel contains a pair of doublets, of Ohara S-FSL5 (S-FSL5Y for the Blue camera) and PBL35Y. They would be bonded with Norland NOA88, which has superb UV transmittance. There is a very close CTE match between these glasses. FEA tests at AAO suggest that a relative change in scale of 50ppm is acceptable, while the largest difference for MSE is several times smaller, even for lenses in the NIR camera. It is not yet proven that the glue works at these temperatures, but there is no reason to suppose otherwise.

The initial design included doublets of Nikon 4786/PBL1Y, as for Hector, with a significant CTE difference (3.6ppm/°C). However, the resulting thermal stresses are unlikely to be acceptable for the much larger temperature changes between assembly and use for MSE, if the slit and optical cameras are all cooled sufficiently to allow use H-band use. S-FSL5/PBL35Y is not as good a combination, so there is an image quality penalty everywhere, driven by the H-band requirement.

The design depends fundamentally on very strong aspheric surfaces. There are aspheres on each face of the first doublet, and on the first face of the second, and on the rear face also for the Blue camera. The aspheric departure is up to 2.4mm. For Hector, with aspheric departures up to 1.2mm, quotes have been obtained and are less than a factor two more expensive than a simple spherical lens. The larger projected fiber size for MSE means that tolerances can be relaxed somewhat; the crucial tolerance being the error in the aspheric slope, specified for Hector as an RMS slope error < 10μrad. The gull-wing nature of L1 means it must be tested in transmission, but this is preferable in any case. The lens sizes are surprisingly modest, with largest aperture 216mm. Image quality can be improved by increasing the lens sizes.

### 3.7 Field-flattener / Dewar window

The ideal material for field flattener use in fast low-dispersion spectrograph cameras should be affordable, easily polished, non-radioactive, high-transmission, high-index and low-dispersion. Extensive lab tests at AAO have so far failed to find such a material. All lanthanum crown glasses tested so far – by Ohara, Schott and Nikon - have variable amounts of alpha particle radiation; this interacts with E2V 231-series detectors to form very strong trails. By far the best glass tested to date is Nikon 7054, an i-line glass with superb transmission, but still giving ~30 such trails per detector per hour.

Several solutions are being explored for Hector. The first is a very thin (0.5mm) fused silica coverslip glued to the inner surface of the field flattener, currently undergoing FEA analysis, but unlikely to be viable for MSE's cooled optics because of the CTE difference with Nikon7054. A sapphire coverslip would be very well matched in CTE, but the birefringreance gives a modest image quality penalty. The second solution is to move to a Fused Silica lens entirely, but the image quality is very badly compromised. A third option is AlON (Aluminium OxyNitride, a clear and fantastically strong ceramic used as transparent armour). The main drawbacks of AlON are a slight haziness costing a few % throughput at all wavelengths, a higher cost for the polishing, and a microroughness limited to ~5-10nm (as compared with the 2nm normally specified for glass lenses, so an order of magnitude more scattered light per surface). However, the scattered light level remains less than any realistic value for the VPH gratings, and is also much closer to focus, meaning a smaller fraction of the scattered light is outside the core of the PSF. The design presented here uses AlON. Other transparent ceramics such as YAG may offer solutions in future.

In all cases, the field-flattener is aspheric on the front concave surface, and planar on the rear surface. The Hector design allows a decentering between field-flattener and detector, but this was not found to give a useful advantage here.

The clearance between Dewar window and detector is just 1mm. This is as specified for Hector, and SpecInst say they can accommodate it in their 1100S series dewars with E2V 231-series detectors. Hawaii4RG15's should be easier, since they have no features protruding forward of the detector surface.

### 3.8 Detectors

Optical detectors are assumed to be E2V 231-series, Back-Illuminated for the Blue arm, Deep Depletion for the Green, and possibly high rho/bulk silicon for the Red. The latter does not yet exist in 231-format, the ROM quotation from E2Vshows that this version of detector is 75% more expensive than standard deep depletion. The NIR arm detector will have to be a Teledyne Hawaii 4RG15.

### 3.9 Detector usage

The projected slit curvature depends on spectral resolution, and hence varies between arms and modes. However, the cameras have pincushion distortion, allowing a certain amount of slit curvature variation to be accommodated, while still keeping all images at all wavelengths on the detectors. Figure 7 shows the layout of the images on the detectors, for all 4 arms in both modes.

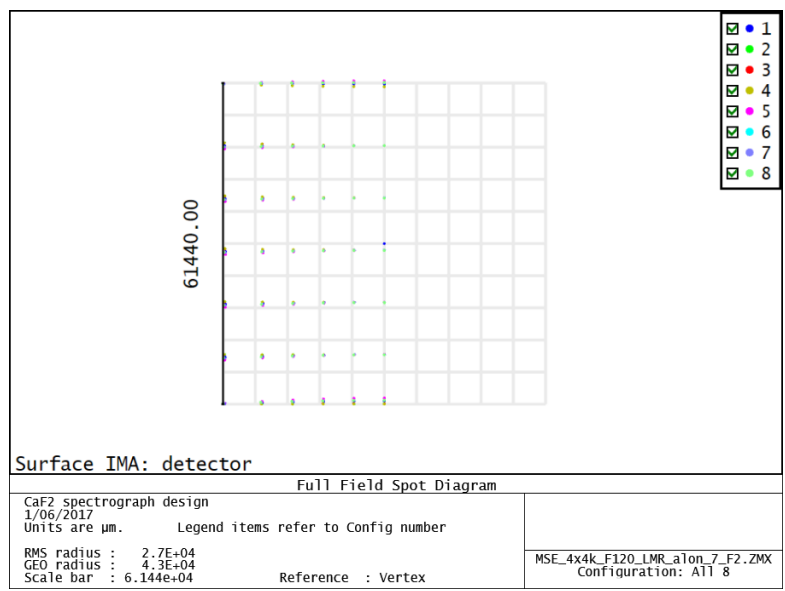

Figure 7. Image layout on the detector for all 4 arms in both modes. Only half the slit is shown. The variation in slit curvature is less than the pincushion distortion in the cameras, and hence causes no loss of detector usage efficiency.

### 3.10 Stops

The formal Stop in this design is virtual, being at the center of curvature of the slit. There is no requirement that the chief rays from each fiber pass through the center of the grating, or any other physical optic. The optics are sized to accept all the light exiting the fibers at F/2.083, and the lens sizes and masses are minimized as part of the optimization. This means that some rays exiting the fibers at faster than F/2.083 can pass through the system too, increasing throughput but harming image quality.

Oval masks will be added to the gratings and cameras to exclude the majority of such rays, and determine the best compromise between throughput and image quality. This exercise has been done for Hector, and shows only a very small change in either, compared with the design with a virtual stop.

### 3.11 Tolerancing and testing

No tolerancing analysis has been done yet. However, given the decreased image quality and increased speed compared with Hector, the tolerances will be very similar. For Hector, the most difficult tolerances are (a) the lens centerings (especially within the doublets), which must be good to 10-20μm; and (b) the surface slope errors, which must be less than ~10μrad RMS. Both are demanding, but within the capabilities of optics vendors specializing in aspheres.

L1 is best tested in transmission, because of the gull-wing asphere. But this is preferable in any case, since it then includes and corrects for inhomogeneity, at least for an on-axis beam. If L1 is made first, then L2 rough polished, then bonded to L1, then the doublet polished and tested in transmission, then errors in the alignment can be polished out, and also the surface errors have some compensation rather than adding in quadrature. The same procedure could be followed for L3/L4.

### 3.12 Image quality

Figure 8 shows the RMS image quality for all configurations. As well as fields along the center of the slit, fields offset by 43μm are shown; the average of the two is close to the average image quality across the fiber. The goal was to maintain an RMS radius less than a quarter of the projected fiber radius (~60μm) for all configurations, wavelengths and slit positions, and this is achieved.

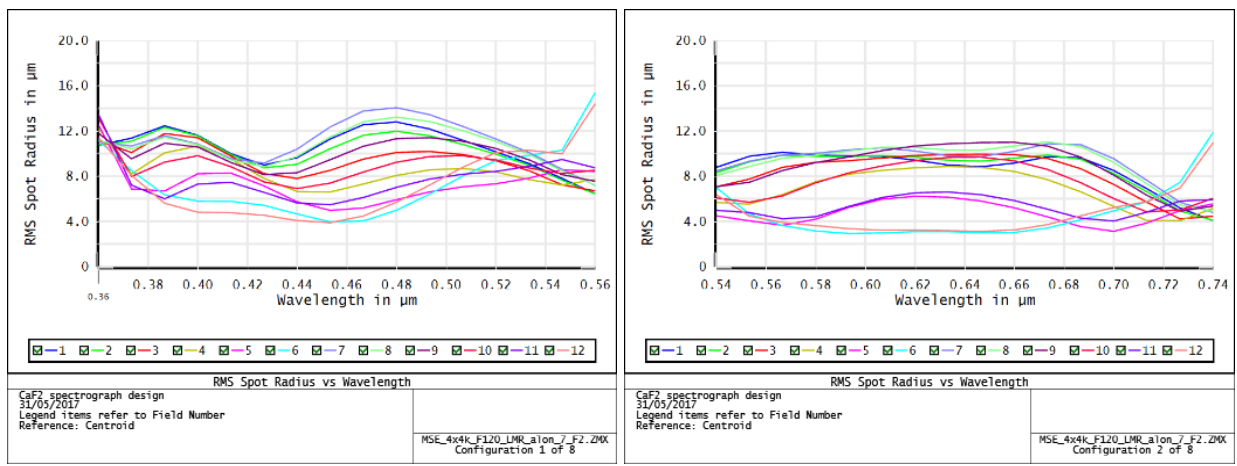

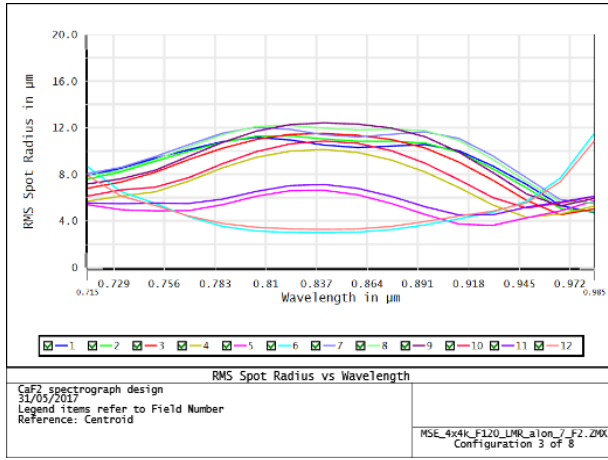
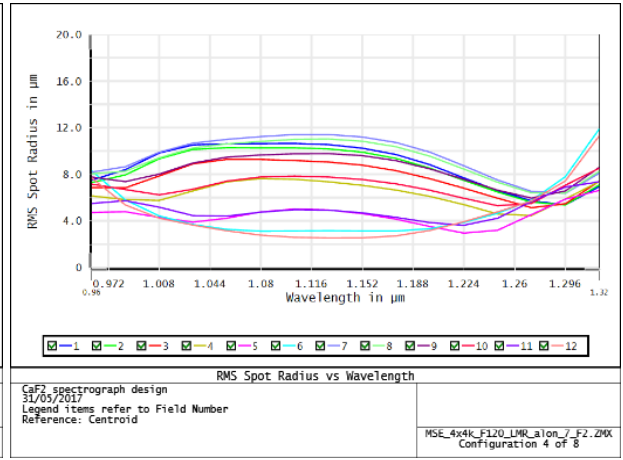
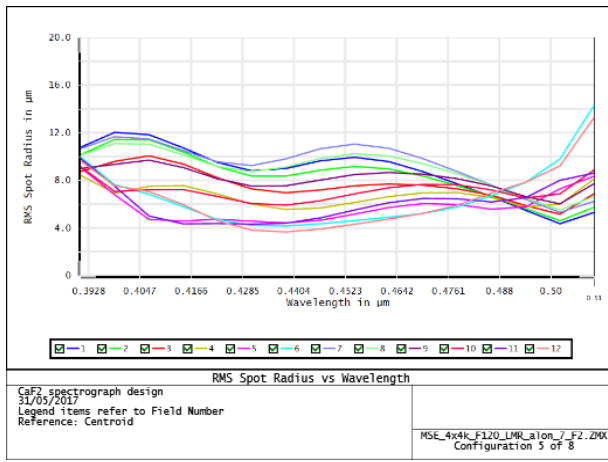
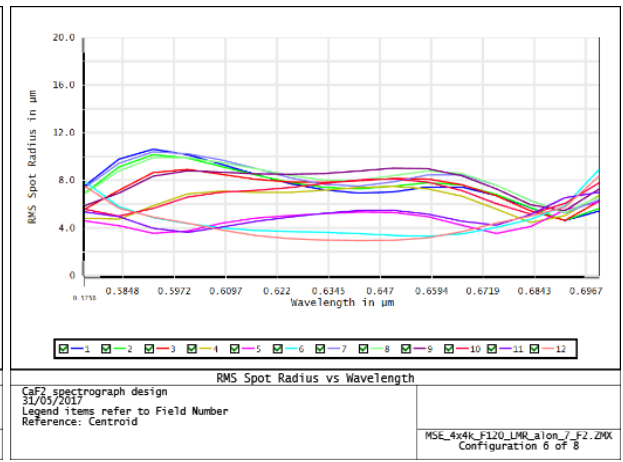
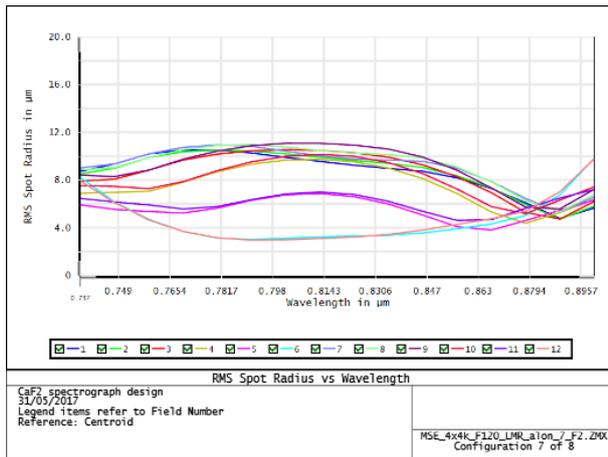
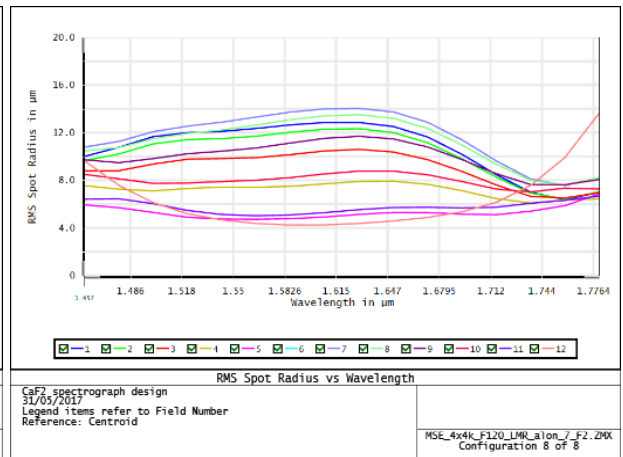

Figure 8. RMS spot radius vs field position and wavelength for each configuration: Blue, Green, Red, NIR cameras, for LR and MR modes.

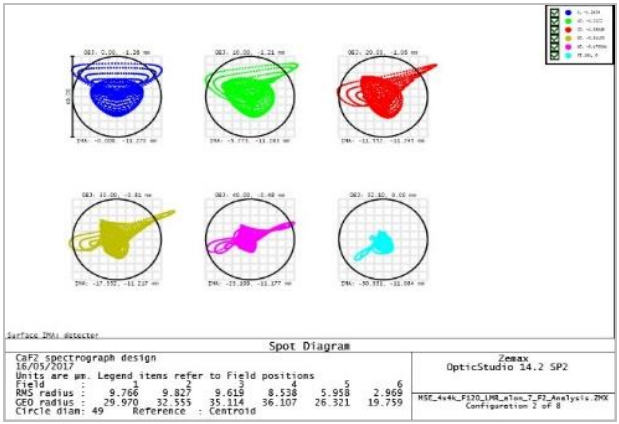
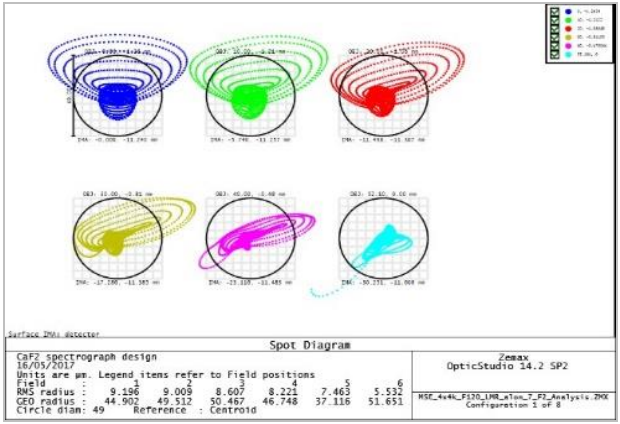
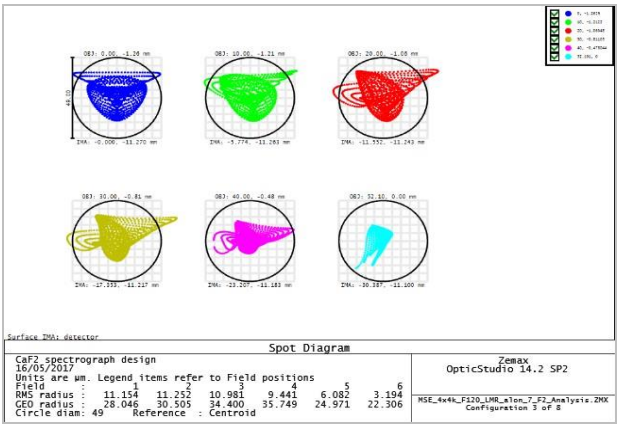
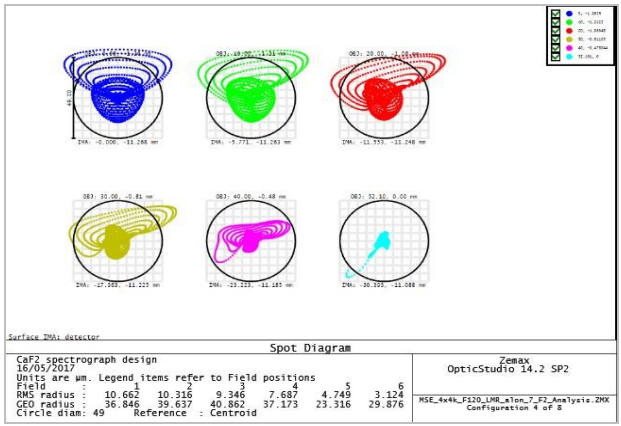
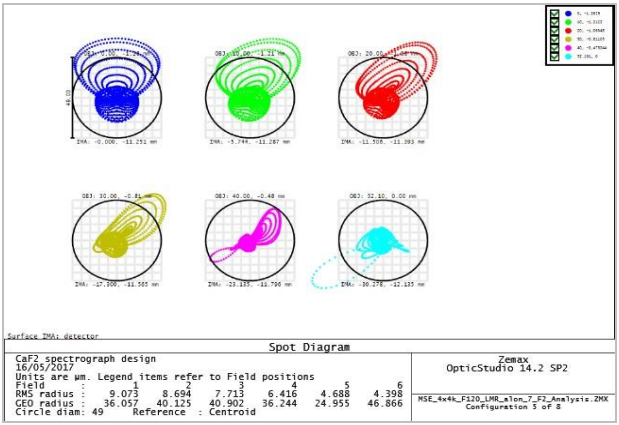
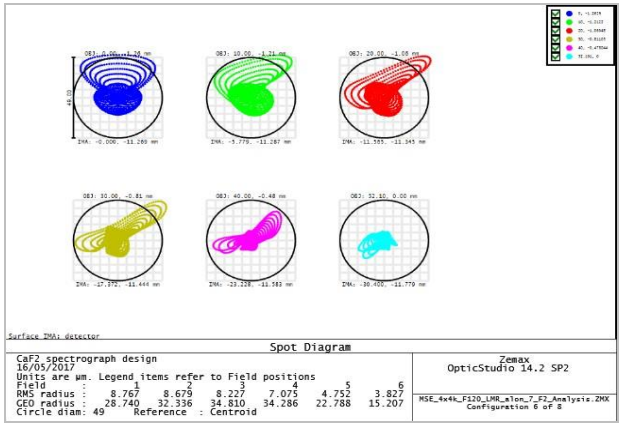

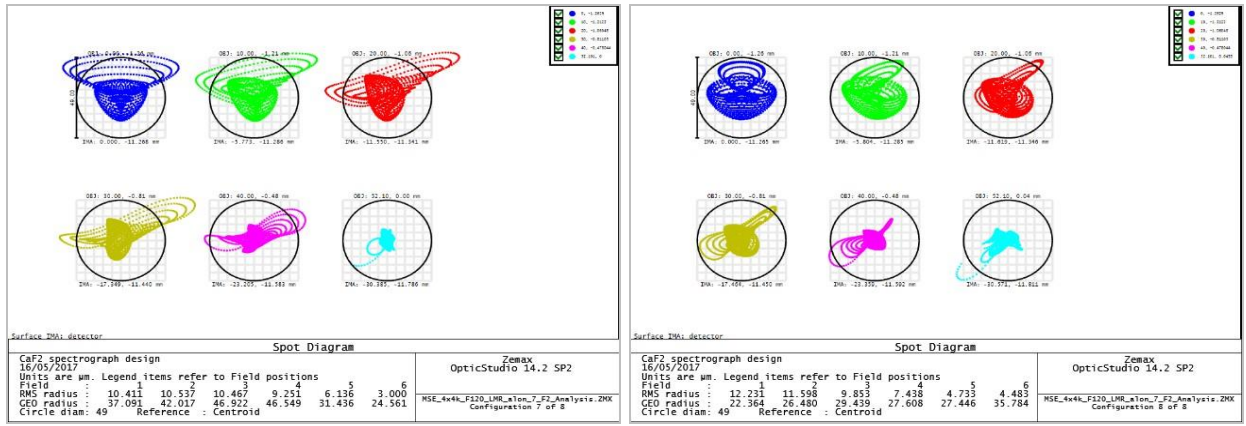

Figure 9. RMS spot diagram vs field position and wavelength for each configuration: Blue, Green, Red, NIR cameras, for LR and MR modes.

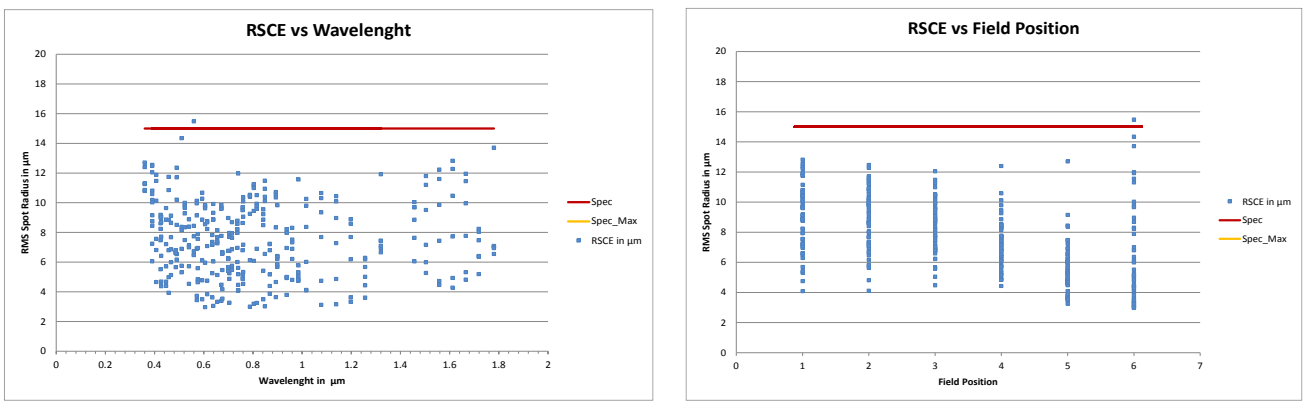

Figure 10. RSCE across Wavelength and field position

### 3.13 Spectral Resolving Power and Spectral sampling Calculation

Following the LMR Requirements Document, Spectral resolution is defined as $R = \lambda / \Delta\lambda$ . The resolution element, $\Delta\lambda$ is the full width at half maximum (FWHM) of a Gaussian function fitted to an unresolved spectral line with wavelength $\lambda$ on the extracted spectrum. Moreover, Spectral sampling is defined as the size in pixels of $\Delta\lambda$, i.e. the width of a resolution element.

The Spectral resolving power is computed as following. The pixellated image is collapsed into a one-dimensional line spread function in the spectral direction by summing rows of five pixels centred on the geometric image. This is illustrated in Figure 11. A Gaussian fit (or else) is made to the line spread function, and the full-width-half-maximum of the fitted curve is obtained. This FWHM defines the resolving element of the spectrograph.

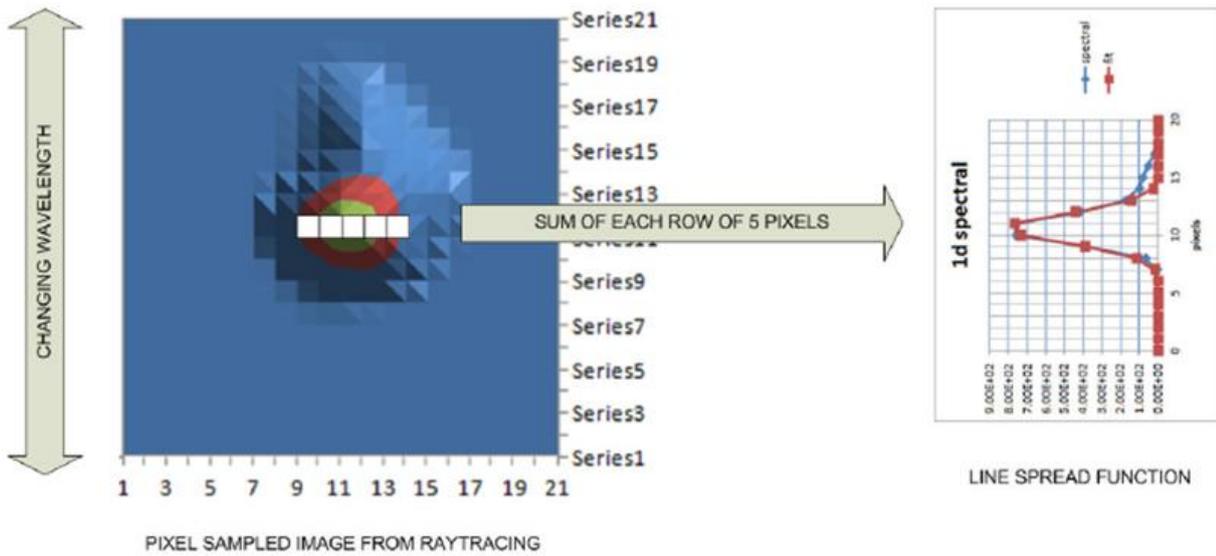

Figure 11: Summing of rows of 5 pixels to give the line spread function

The spectral and spatial sampling are first defined by the fiber image dimension. As shown below this image is distorted by the optical system and more particularly in the spectral direction because of the dispersion grating. Fiber image dimensions therefore vary in field position and wavelength:

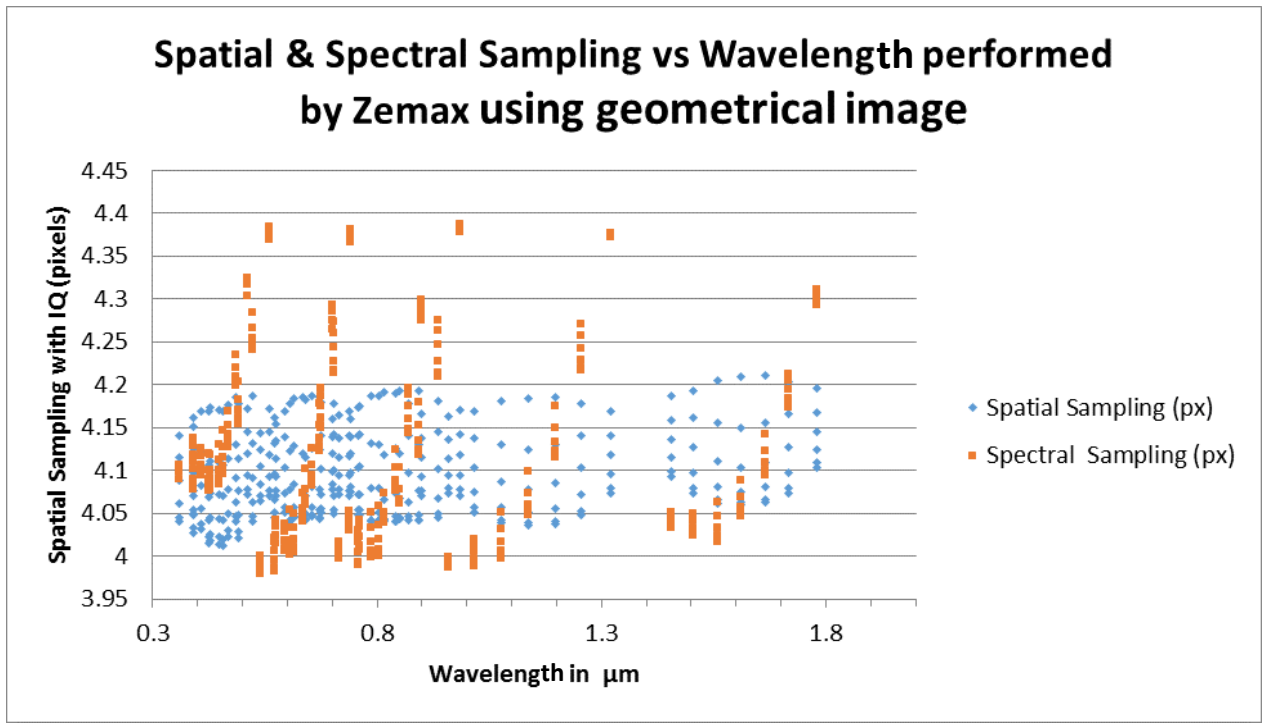

Figure 12: Spatial & Spectral geometrical sampling

From this initial geometrical spectral sampling different effects are to be taken into account to calculate the final resolution:
- Fiber image degradation caused by optical image quality, CCD diffusion, stray light, …
- Image collapse → Line Spread Function (LSF)
- LSF fit → FWHM in pixel

We developed for 4MOST a software enabling those operations to be done taking as input:
- Zemax geometrical image size
- Zemax calculated PSF
- CCD deep depletion diffusion curve (especially for the visible blue range)
- Internal scattered light Lambertian model
- Global defocus related to CCD flatness

One of the conclusions of our first iteration on Gaussian fit was that it was too optimistic, the project circle assuming no image quality degradation was on the other hand too pessimistic. We therefore proposed an alternative custom fit method

In this custom model, the Line Spread Function is modelized using a filled circle, mimicking the fibre convolved by a PSF modelled as a circular 2D Gaussian (parametrised by its FWHM). This convolved circle is rebinned (resampled) using the detector pixel size and summed up over 5 pixels in the spatial direction. The best model is found by minimizing the difference between this model and the raw data (also summed over 5 pixels in spatial direction) using a least squares algorithm. Currently the raw data are provided by the optical PSF derived from Zemax, convolved by a mock fibre (as for the model), rebinned onto detector pixels and summed up over 5 pixels (Figure 13)

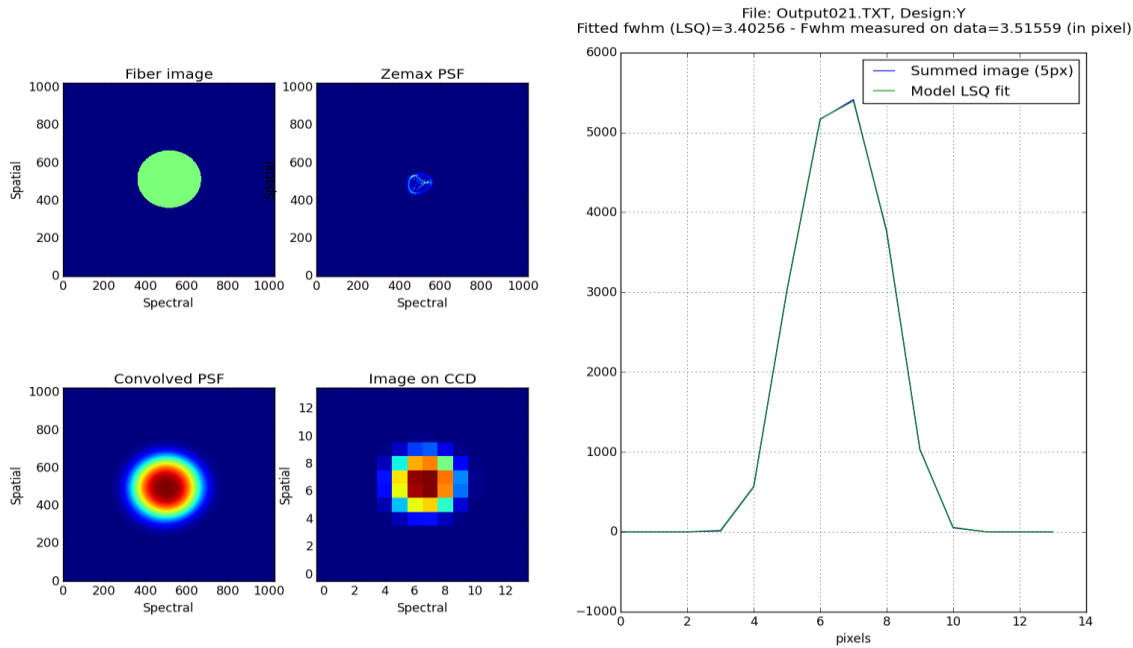

Figure 13: Custom model. (Left) simulated raw fibre image: the circular fibre (upper-left panel) is convolved by the theoretical PSF from Zemax (upper-right panel), then projected onto the CCD pixel grid (bottom right panel). (Right) the summed image over the central 5 detector columns (blue curve) is fit by the custom model (green curve)

As a matter of comparison, please find below the unbinned profil of the summed image compared to the 3 possible fits, Gaussian, Projected circle and custom model. This calculation has been done on the average field position (20 mm from center) at λ = 457 nm with the LMR CoDR Reference Zemax optical design.

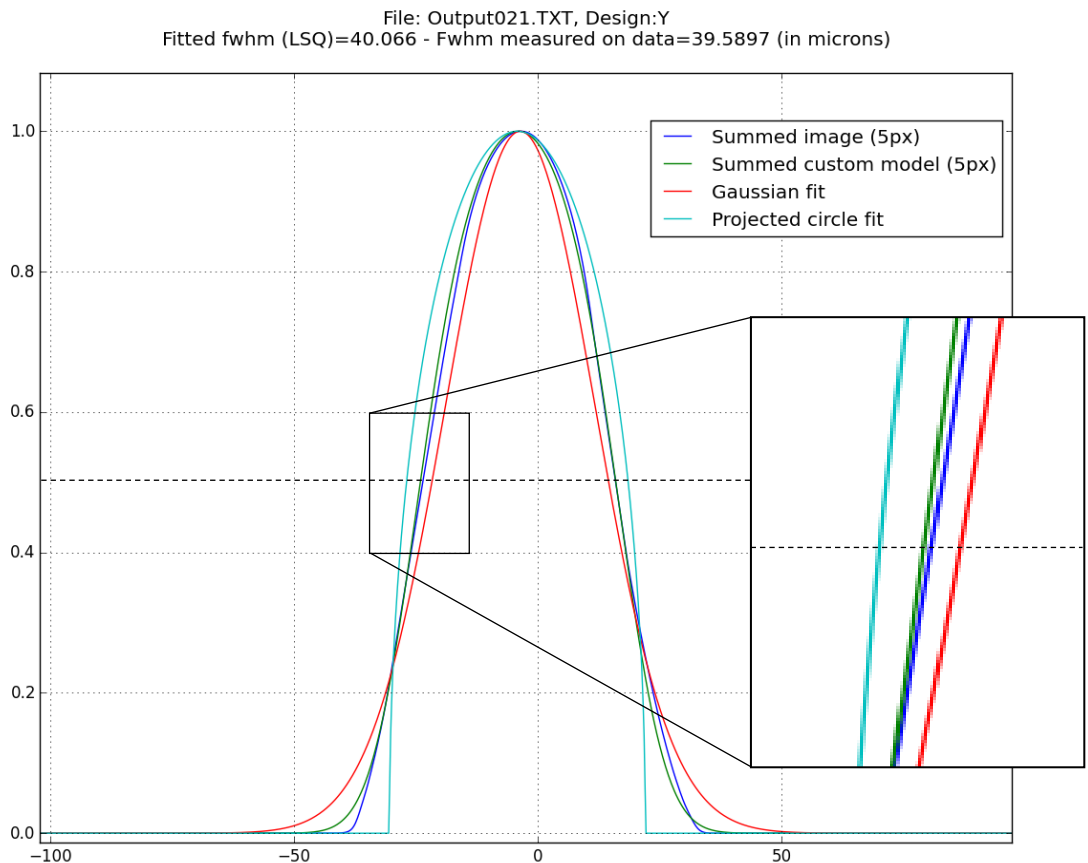

Figure 14: Cut derived from a summed image over 5 pixels, a summed custom model, a Gaussian fit and a projected circle fit

As the custom is more accurate and is conservative with regards to the Gaussian fit, it has been chosen to present the custom fit results.

### 3.14 Spectral Sampling

As calculated with the custom model, the spectral sampling results are as following:

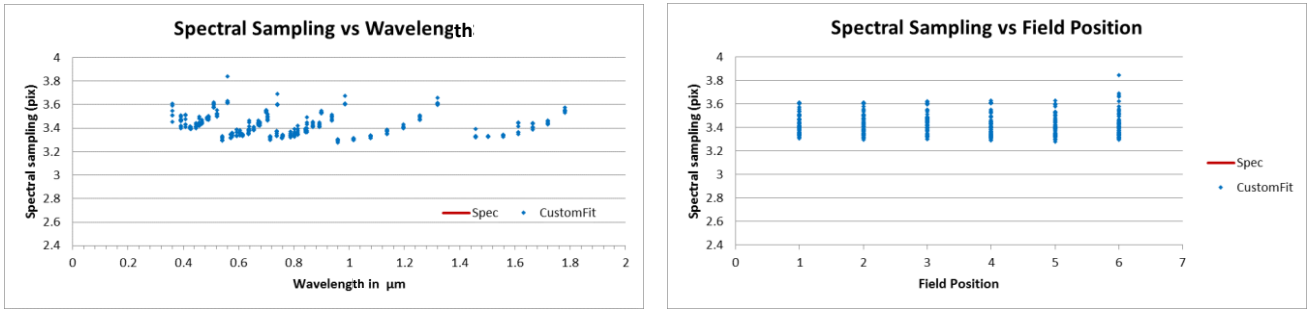

Figure 15: Spectral Sampling according wavelength and field position

### 3.15 Spectral Resolution

Associated with the above spectral sampling the calculation of the spectral resolution is straightforward. As reminder the result of this calculation includes many optical effects described in §3.13. It does not however include fabrication and alignment tolerance effect.

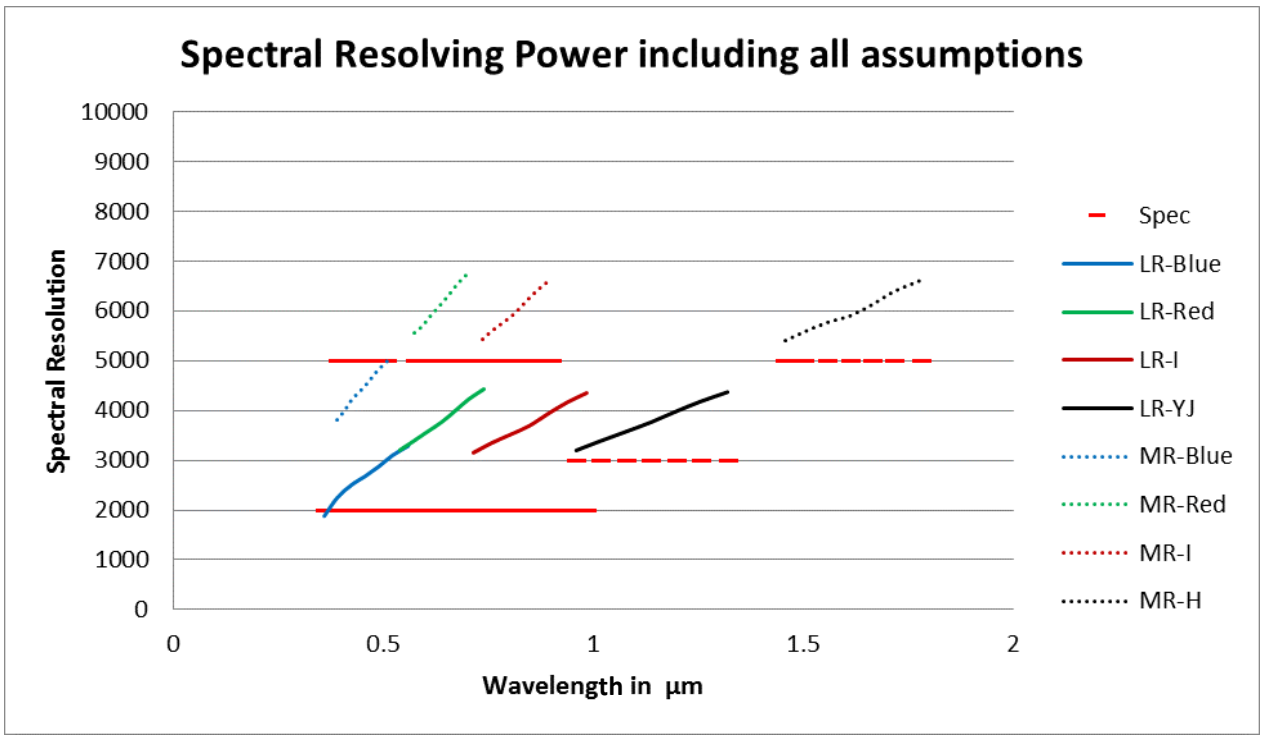

Figure 16: Spectral Sampling variation according wavelength

As one can see the spectral resolution is not met at all the in the moderate blue arm. This feature is inherited from the Hector design which benefits then in the green and red arm of an increased resolution. As shown in §3.1 a rebalancing of the wavelength range is possible but the holding of the 5000 minimum resolution in moderate resolution is very challenging without moveable arms. The reason for that is that the spectral coverage and dispersive power in moderate resolution are directly related to the Sapphire index.

### 3.16 Crosstalk

The crosstalk between fibers will happen inevitably as fibers are close to each other in the Field of View and the optical image quality and scattered light will send light away from the fiber image. The crosstalk calculation is done with the same software and optical effect described in §3.13. The crosstalk criterion is derived by computing the ratio between the flux contained in a box centered on a monochromatic LSF of 3x3 pixel (this corresponds more or less to the spectral sampling) and the same box separated by 6.8 pixels (corresponding to 170 µm step in the fiber slit) on the spatial direction where the neighboring fiber spectrum is to be read. The results are the following:

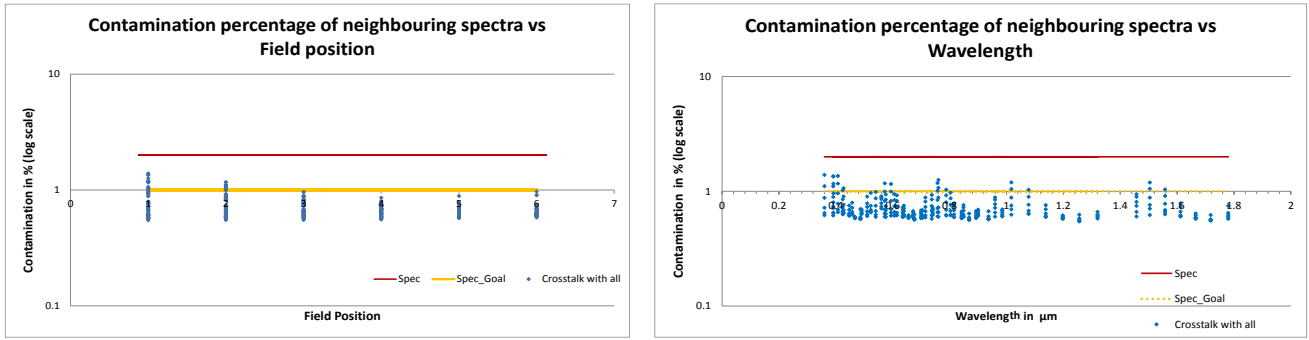

Figure 17: Fiber to fiber crosstalk according to field position and wavelength

As stated crosstalk criteria is met for all points on all detectors. The 1% Goal is met for 93% of them

### 3.17 Throughput

Figure 18 shows the theoretical performance of the MSE LR gratings. In each case, peak efficiency is ~93%, minimum efficiency ~72%, and average efficiency 86%. Real performance should only be ~1% worse than this. The MR gratings will be somewhat less efficient (the larger grating angle increasing the polarization splitting), and values of 87.5% peak, 65% minimum, and 80% average have been assumed.

Coatings are assumed to be simple 3-layer hard coatings on all optics below the dichroics, and solgel+MgF2 on the slit lens. Dichoics are assumed to be 97.5% efficient in both reflection and transmission. Internal transmission is calculated at the center of each lens, except for the field-flattener, where an average value of 5mm is assumed. Detector efficiencies are taken from E2V or Teledyne. Collimator overfilling losses are included and assumed to be 5%. The collimator mirror is assumed to be enhanced silver coated.

In LR mode, overall throughput is 42% at 360nm, and exceeds 50% for 370-1800nm, with average better than 60%. In MR mode, overall throughput exceeds 45% everywhere, with average better than 55%.

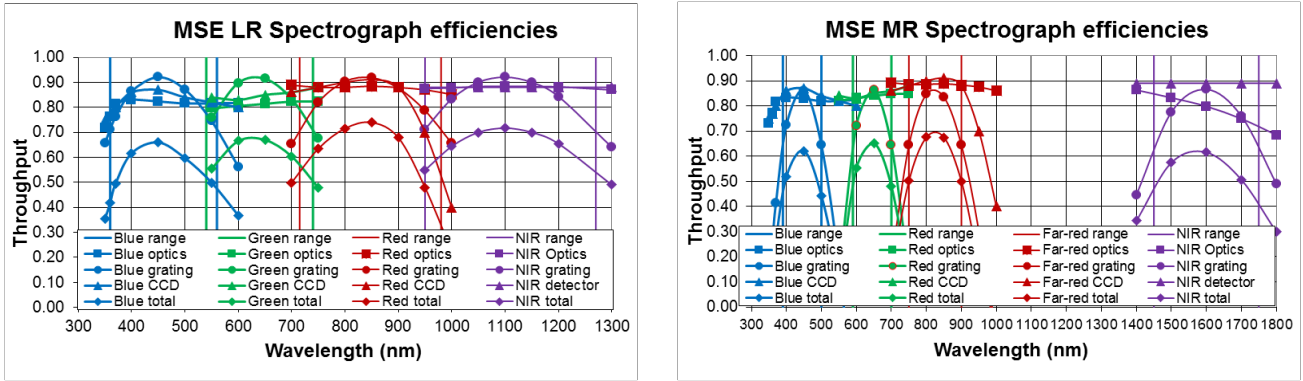

Figure 18. Throughputs for LR & MR mode.

Though everything is shown compliant here, one has to notice than the requirement of 50% is barely met on the average Moderate resolution. Please note as well that fabrication tolerance may influence negatively the figure above.

## 4 CONCLUSION

As detailed in this paper we see that the main non-compliance of the LMR spectrograph concept is identified on the spectral resolution. A new design is on the way to improve this. Some partial compliance are also worth being discussed but don't seem absolutely critical. Some uncertainties remain but the most important one is certainly the thermal background aspect and possible evaluation of that. This should be one of the main task for the upcoming phase.

Finally I would like to thank, Will Saunders for his great LMR optical design and his optical description section of this document, Jean-Emmanuel Migniau for the LMR mechanical design, Florence Laurent and Arlette Pecontal for doing all the custom software optical calculation and associated analysis, Pierre-Henri Carton and Christophe Yèche for their inputs on the detector system and Johan Richard for his scientific support.